\documentstyle[12pt]{article}

\begin{document}

\begin{center}

{\large \bf The Binary Pulsar Is Not the Ultimate Test for the 
Theory of Gravity } \\

\vspace{2.0cm}

M. Novello and V. A. De Lorenci \\
\vspace{0.6cm}
{\it Laborat\'orio de Cosmologia e F\'{\i}sica Experimental
de Altas Energias.} \\
\vspace{0.1cm}
{\it Centro Brasileiro de Pesquisas F\'{\i}sicas,} \\
\vspace{0.1cm}
{\it Rua Dr. Xavier Sigaud, 150, Urca} \\
\vspace{0.1cm}
{\it CEP 22290-180 -- Rio de Janeiro -- RJ. Brazil.} \\
\vspace{2.0cm}

\begin{abstract}

General relativity can be formulated either as in its original 
geometrical version (Einstein, 1915) or as a field theory (Feynman, 1962). 
In the Feynman presentation of Einstein theory an hypothesis 
concerning the interaction of gravity to gravity, which was hidden in the 
original version, becomes explicit. This is nothing but the assumed 
extension of the validity of the equivalence principle not only for 
matter-gravity interaction, but also for gravity-gravity. 
Recently we have presented a field theory of gravity (from here on called 
the NDL theory) which does not contain such a hypothesis.
We have shown that, for this theory, both the cosmological structure and the 
PPN approximation for the solar tests are satisfied. 
 
The proposal of this paper is to go one step further and to show that NDL 
theory is able to solve the problem of radiation emission by a binary pulsar 
in the same degree of accuracy as it was done in the GR theory. In the 
post-Newtonian order of approximation we show that the quadrupole formula of 
this theory is equal to the corresponding one in general relativity.
Thus, the unique actual observable distinction of these theories concerns 
the velocity of gravitational waves, which becomes then the true 
ultimate test for gravity theory.

\end{abstract}

\end{center}

\vspace{1cm}

Pacs numbers: 04.20.Cv, \, 04.80.Cc

\newpage

\section{Introduction}
\label{introduction}

\subsection{Introductory Remarks}
\label{remarks}

The general relativity (GR) description of the observations concerning the 
rate of the energy lost by the binary pulsar 
increased enormously the status of this theory. Thus, any theory that
dares to be competitive with GR should at least be able to
provide the same degree of accuracy in the explanation of this 
phenomenon (besides, of course, all remaining standard observational tests, 
which means PPN and cosmology). 

Although GR is presented as an universal modification of the metrical 
properties
of spacetime, an alternative way to describe GR as a field theory in the 
same lines as any other interaction was revived by some authors
\cite{Deser,GPP}. The idea goes back to Feynman\footnote{The original
issues seems to be firstly worked out by Gupta \cite{Gup54} and Kraichnan 
\cite{Krai55}.} investigation 
\cite{Feynman}. Indeed, in his 1962 {\it Lectures on Gravitation} it has been
shown that a field theoretical approach of gravity should be
possible and its basic ingredients should deal, besides the field 
$\varphi_{\mu\nu}$ itself, with two 
metric tensors: an auxiliary one $\gamma_{\mu\nu}$ --- which is 
not observable --- and an effective one $g_{\mu\nu}$ related by 
$g_{\mu\nu} = \gamma_{\mu\nu} + \varphi_{\mu\nu}$.

The basic hypothesis of GR concerns the extension of the 
equivalence principle beyond its original domain of experimental evidence, 
that concerns material substance of any form, the adoption of its validity not only by matter or
non-gravitational energy of any sort but also by gravity energy itself. 
Such an universality of interaction is precisely the cornerstone that
makes possible the identification of a unique overall geometry of spacetime 
$g_{\mu\nu}$. In Eintein GR the properties of gravity are associated
to the Riemannian curvature, which becomes then the equivalent substitute of 
gravitational {\it forces}. We remark that, although such geometrization
scheme is permissible, it is by no way mandatory. All observable characteristics 
and properties of Einstein theory can be well described in terms of a 
field $\varphi_{\mu\nu}$. Indeed, the lesson we 
learn from Feynman approach is this: {\bf the geometrical description of GR 
is nothing but a choice of representation}. Let us
emphasize that such an alternative description of GR in no way sets a
restriction on it, but only enlarges its power of understanding.

From this approach it follows that contrary to a widespread belief, 
GR can be described in terms of a two metric
structure\footnote{This is also the case in NDL theory.}. 
Furthermore, Feynman has shown that the coherence of a
spin-2 theory that starts with the linear Fierz-Pauli \cite{Fierz} equation 
written in terms of the symmetric field $\varphi_{\mu\nu}$ in a Minkowskian
spacetime requires, in a very natural way, due to the self-interaction process 
described above, the use of an induced metric tensor, the quantity 
$g_{\mu\nu}$. Thus, as we announced above, the field-theoretical way of 
treating GR appeals to 
a two metric structure. This is the {\it standard} procedure. Nevertheless, 
and just by tradition, this is not the way that Einstein 
theory is presented in the text books\footnote{The absence 
in the literature of such alternative but equivalent way to present Einstein 
theory of gravity seems to be the main responsible for the young students of 
theoretical physics to understand GR as a completely separate and 
different theory from any other field.}. This interpretation allows us to 
state that two-metric theories of gravity are less exotic than it is 
usually displayed \cite{Willbook}. Let us emphasize that the second 
metric is nothing but a 
convenient auxiliary tool of the theory. It is not observable 
(neither in general relativity nor in NDL theory) and as 
such can be eliminated from a description made only in terms of observable 
quantities. In textbook presentations of GR one makes the choice of 
a unique geometry. This, of course, does not preclude an
alternative equivalent description \cite{GPP}.

Recently we have exploited some consequences of such field theoretical
description of gravity adding a new ingredient: we do not require the
extrapolation to gravitational energy of the 
hypothesis of universality of the equivalence principle (EP) 
although, as we shall see, it contains many 
of the ingredients of GR. The main lines of NDL theory can be 
synthesized in the following statements:

\begin{itemize}
 \item{Gravity is described by a symmetric second order tensor 
$\varphi_{\mu\nu}$ that satisfies a non-linear equation of motion;}
 \item{Matter couples to gravity in an universal way. In this interaction, 
the gravitational field appears only in the combination 
$g_{\mu\nu} = \gamma_{\mu\nu} + \varphi_{\mu\nu}$. Such tensor 
$g_{\mu\nu}$ acts as the true metric tensor of the spacetime as seen by matter 
or energy of any form except gravitational energy;}
 \item{The self interaction terms of the gravitational field break the 
universal modification of the spacetime geometry.}
\end{itemize}

It has been conjectured \cite{Willbook} that the observation of the binary 
pulsar should
be the ultimate test of gravity theory. This is due to the fact that so far 
the alternative models that have been proposed to explain gravitational 
processes did not succeeded in provide coherent explanation of observations, 
mainly concerning the pulsar PSR $1913 + 16$.

However this statement seems to be not true. 
Indeed we shall prove in the present paper that NDL theory is able to provide 
a description of the gravitational radiation emitted by the pulsar.

\subsection{Synopsis}
\label{synopsis}

In Section \ref{notacao} we introduce the definitions and symbols 
we are using.

In Section \ref{ShortGR} we summarize the standard Gupta-Feynman-Deser
(GFD) approach for the field theory of gravitation that led to general relativity.

In Section \ref{ShortNDL} we make a short review 
of our previous paper \cite{NDL} and compare with GFD approach.
We show the main features of NDL theory with emphasis on the propagation of 
gravitational waves.

In Section \ref{Quadrupole} we present a short derivation of the
gravitational quadrupole formula to radiation emission by a 
binary system. We compare then our results with the GR.

We conclude in Section \ref{Conclusion} with some comments concerning
both theories, mainly with respect to the classical test and
binary pulsar. We end with some comments and perspectives for 
future investigations.

\section{Definitions and Notations}
\label{notacao}

In order to exhibit the complete covariance of the theory 
all quantities will be described in an arbitrary system of
coordinates. In the auxiliary background geometry
of Minkowski spacetime of metric $\gamma_{\mu\nu}$ the 
covariant derivative, represented by a semi-comma, is 
\begin{equation}
V_{\mu ;\nu} = V_{\mu ,\nu} - \Delta^{\alpha}_{\mu\nu} V_{\alpha}
\label{d0}
\end{equation}
in which the associated Christoffel symbol 
$\Delta^{\alpha}_{\mu\nu}$ is given by
\begin{equation}
\Delta^{\alpha}_{\mu\nu} = 
\frac{1}2 \gamma^{\alpha\beta}\hspace{0.1cm}( 
\gamma_{\beta\mu ,\nu} + 
 \gamma_{\beta\nu ,\mu} - \gamma_{\mu\nu ,\beta}).
\label{d01}
\end{equation}
The corresponding curvature tensor vanishes identically, that is
\begin{equation}
R_{\alpha\beta\mu\nu}(\gamma_{\epsilon\lambda}) = 0.
\label{d02}
\end{equation}

We define a three-index tensor $F_{\alpha\beta\mu}$, which we will call 
the {\bf gravitational field}, in terms of the symmetric standard variable 
$\varphi_{\mu\nu}$ (which will be treated as the potential) to describe 
spin-two fields, by the expression
\begin{equation}
F_{\alpha\beta\mu} = \frac{1}{2} ( \varphi_{\mu[\alpha;\beta]} + 
F_{[\alpha}\gamma_{\beta]\mu} )
\label{d1}
\end{equation}
where we are using the anti-symmetrization symbol $[\hspace{0.9mm}]$ like 
\begin{equation}
[x, y] \equiv xy - yx.
\label{d4}
\end{equation}
We use an analogous form for the symmetrization symbol $(\hspace{0.9mm})$ 
\begin{equation}
(x, y) \equiv xy + yx.
\label{d5}
\end{equation}
The quantity $F_{\alpha}$ is the trace 
\begin{equation} 
F_{\alpha} = F_{\alpha\mu\nu} \gamma^{\mu\nu} 
\end{equation}  
that is,
\begin{equation} 
F_{\alpha} = \varphi_{,\alpha} - \varphi_{\alpha}\mbox{}^{\epsilon}
\mbox{}_{;\epsilon}.
\end{equation}

From the above definition it follows that this quantity $F_{\alpha\beta\mu}$ 
is anti-symmetric in the first pair of 
indices and obeys the cyclic identity, that is
\begin{equation}
F_{\alpha\mu\nu} + F_{\mu\alpha\nu} = 0,
\label{d2}
\end{equation}
and
\begin{equation}
F_{\alpha\mu\nu} + F_{\mu\nu\alpha} + F_{\nu\alpha\mu} = 0.
\label{d3}
\end{equation}

From the field variables we can form the invariants\footnote{Note 
that, besides this invariants, it is possible to define a quantity $C$, constructed
with the dual, that is $C \equiv F^{*}_{\alpha\mu\nu}\, F^{\alpha\mu\nu}.$
We will not deal with such quantity here.}:
\begin{eqnarray} 
A &\equiv& F_{\alpha\mu\nu}\hspace{0.5mm} F^{\alpha\mu\nu},
\nonumber \\
B &\equiv& F_{\mu}\hspace{0.5mm} F^{\mu}. 
\label{AB}
\end{eqnarray}

Greek indices run into the set \{$0,1,2,3$\}, while Latin indices run
into the set \{$1,2,3$\}. 
Finally, the quantity $\kappa$ is related to Newton\rq s constant 
$G_{\mbox{\tiny N}}$ and the velocity of light $c$ by the definition 
\begin{equation} 
\kappa = \frac{16\pi G_{\mbox{\tiny N}}}{c^{4}}. 
\end{equation} 
We set $G_{\mbox{\tiny N}}=1$ and $c=1$.

\section{The GR Theory of Gravity: A Short Resume}
\label{ShortGR}

General relativity takes for granted that gravity is nothing but the fact
that all existing form of energy/matter 
interacts through the modification of the universal geometry. However, 
such a view is not exclusive and it is conceivable to try 
to use two metrics to describe in an equivalent 
way all content of such theory. There is no simpler and more direct way to 
prove this statement than the one set forth by Feynman. It is worth
to remark that a such duplication 
causes no further difficulties when one realizes that the second 
auxiliary metric $\gamma_{\mu\nu}$ is unobservable. 

Let us pause for a while and make, just for completeness, a summary of the 
principal features of this equivalent scheme. 
The theory starts with the Fierz-Pauli linear equation
\begin{equation}
G^{\mbox{\tiny (L)}}_{\mu\nu} = - \, \kappa\,T_{\mu\nu} 
\label{111}
\end{equation}
in which $T_{\mu\nu}$ is the matter energy-momentum tensor and 
$G^{\mbox{\tiny (L)}}_{\mu\nu}$ is a linear operator defined by:
\begin{equation} 
G^{\mbox{\tiny (L)}}_{\mu\nu}\equiv \Box\phi_{\mu\nu} -
\phi_{\ \mu ;\alpha\nu}^{\alpha}
-\phi_{\ \nu ;\alpha\mu}^{\alpha} +\phi_{\ \alpha ;\mu\nu}^{\alpha} 
-\gamma_{\mu\nu}(\Box\phi_{\ \alpha}^{\alpha} 
-\phi^{\alpha\beta}_{\ \ ;\alpha\beta}). 
\end{equation}

The action for this linear theory is given in terms of the invariants of the
field $F_{\alpha\mu\nu}$ --- defined by Eq. (\ref{d1}) --- that is:
\begin{equation} 
S^{\mbox{\tiny (L)}} =  \int {\rm d}^{4}x\,  
\sqrt{-\gamma} \hspace{0.8mm}( A -  B).
\end{equation}

Since $G^{\mbox{\tiny (L)}}_{\mu\nu}$ is divergence-free it follows for 
coherence that 
the matter energy momentum tensor should also be divergence-free. However
this is in contradiction with the fact that gravity may exchange
energy with matter. To overcome such situation, one introduces an object 
which we call {\it Gupta-Feynman gravitational energy tensor} $ 
t^{(g)}_{\mu\nu} $
--- a cumbersome non linear expression in terms of $\varphi_{\alpha\beta}$
and its derivatives --- that is to be added to the right hand side of 
Eq. (\ref{111}) in order to obtain a compatible set of equations:
\begin{equation}
G^{\mbox{\tiny (L)}}_{\mu\nu} = - \, \kappa\,\left[ t^{(g)}_{\mu\nu} 
+ T_{\mu\nu}\right].
\label{112}
\end{equation}

Note that, instead of using the standard procedure (as it happens in others 
nonlinear theories) --- which in the case we examine here, asks for
the introduction of a nonlinear functional of the invariants $ A $ and $ B $, 
dealt with in the linear case --- in order to obtain the dynamics of GR, 
one must use other functionals of the basic field 
$\varphi_{\mu\nu}$ which are not present in the linear case, 
that means, they are not displayed in terms of the invariants $A$ and $B$. 
We do not intend to repeat here the whole procedure\footnote{The reader may 
consult the references \cite{Deser} and \cite{GPP} for more details.}, 
but only to call the reader\rq s attention to such an unusual treatment of 
dealing with a nonlinear process. 
The origin of this approach goes back to the hypothesis 
of the validity of the equivalence principle for gravitational energy. In the 
next section, we will show that NDL theory follows a more traditional way of 
generalization to a nonlinear theory by the assumption of nonlinear 
functional of the basic invariants dealt with in the linear case. 
Why does GR break this symmetry? What is its motivation? The answer to this
we can find by the assumption of the general validity of the equivalence
principle for all forms of energy, including gravity.

In general relativity since the identification of gravitational processes
to the modified geometry is postulated {\it a priori} there is no room
for the suspection of the assumption of such hypothesis. It is only in its 
Feynman version that it appears netly.
 
Now it is straightforward to show that, for a convenient 
choice of the expression of the {\it Gupta-Feynman gravitational energy}, the 
equation (\ref{112}) is nothing but Einstein dynamics. The main 
steps can be synthesized as follows. 

From the equivalence principle, the observable geometry is given by the
quantity\footnote{In this section we follow the convention as in 
\cite{GPP}. We could use instead $g^{\mu\nu} = \gamma^{\mu\nu} +
\varphi^{\mu\nu}$ or
$g_{\mu\nu} = \gamma_{\mu\nu} +\varphi_{\mu\nu}$, which provide non-equivalent
theories.} 
\begin{equation}
\sqrt{-g}g^{\mu\nu} = \sqrt{-\gamma}\left(\gamma^{\mu\nu} +
\varphi^{\mu\nu}\right).
\end{equation}
Define the tensor 
$K^{\alpha}\mbox{}_{\mu\nu}$ as\footnote{We remind the reader of our convention
that the semi-comma is the covariant derivative with respect to 
Minkowski metric in an arbitrary coordinate system.}:
\begin{equation}
K^{\alpha}\mbox{}_{\mu\nu} = \frac{1}{2} 
g^{\alpha\epsilon}\left[g_{\mu\epsilon;\nu} + g_{\nu\epsilon;\mu}
-g_{\mu\nu;\epsilon}\right],
\label{ka}
\end{equation}
and obtain the Ricci contracted curvature tensor as:
\begin{equation} 
R_{\mu\nu} = -\frac{1}{2}K_{( \mu\,;\nu )} + K^{\alpha}\mbox{}_{\mu\nu\,;\alpha} 
- K^{\alpha}\mbox{}_{\mu\lambda} \,  K^{\lambda}\mbox{}_{\nu\,\alpha} 
+ K^{\lambda}\mbox{}_{\mu\nu} \,  K_{\lambda}. 
\end{equation} 
At this point one has to make a definite choice for $t^{(g)}_{\mu\nu}$
in terms of the quantities $\varphi_{\mu\nu}$ and $K^{\alpha}\mbox{}_{\mu\nu}$.
A rather long but tedious calculation shows that in order to arrive at
Eisntein\rq s equations of motion one must choose (see for instance GPP \cite{GPP})
\begin{equation}
\kappa\,t^{(g)}_{\mu\nu} = -(KK)_{\mu\nu} + \frac{1}{2}
\gamma_{\mu\nu}(KK)^{\alpha}\mbox{}_{\alpha} 
+ Q^{\lambda}\mbox{}_{\mu\nu;\lambda} 
\label{energyGPP}
\end{equation}
in which 
\begin{equation}
(KK)_{\mu\nu} \equiv K^{\alpha}\mbox{}_{\mu\nu}K_{\alpha} -
K^{\alpha}_{\mu\beta}K^{\beta}\mbox{}_{\nu\alpha}
\end{equation}
and 
\begin{eqnarray}
Q^{\lambda}\mbox{}_{\mu\nu} &\equiv& \frac{1}{2}\left\{
  - \gamma_{\mu\nu}\varphi^{\alpha\beta}K^{\lambda}\mbox{}_{\alpha\beta}
  + \varphi_{\mu\nu}K^{\lambda} - \varphi_{(\mu}\mbox{}^{\lambda}K_{\nu)}
  + \varphi^{\beta\lambda}K^{\alpha}\mbox{}_{\beta (\mu}\gamma_{\nu )\alpha}\right.
\nonumber\\
  && + \left. \varphi_{( \mu}\mbox{}^{\beta}\left[ K^{\lambda}\mbox{}_{\nu )\beta}
  - \gamma_{\nu )\alpha}\gamma^{\sigma\lambda}K^{\alpha}\mbox{}_{\beta\sigma}\right]
    \right\}.
\end{eqnarray}

Using the expression (\ref{energyGPP}) into the formula (\ref{112}) one 
obtains finally
\begin{equation}
R_{\mu\nu} -\frac{1}{2}R \, g_{\mu\nu} = -\frac{\kappa}{2} T_{\mu\nu}.
\label{Eintein}
\end{equation}
Note that this expression assures the validity of 
equivalence principle not only for all matter and energy, but
also by gravitational energy.

We can synthesize, this procedure by the statement: 
\begin{itemize}
\item
The interaction of matter and the gravitational energy is nothing but
a universal modification of spacetime geometry.
\end{itemize}

Could it be possible to follow another path to deal with a
nonlinear extension of Fierz original model? The answer is yes and 
led us to the NDL model. Let us see how this can be made in a 
straightforward way.

\section{The NDL Theory of Gravity: A Short Resume}
\label{ShortNDL}

The NDL theory starts at the same point as GR, that is, Fierz linear theory 
for spin two field. However, instead of breaking the symmetry
displayed in the linear regime, presented in the combination of the invariants 
under the form $ A - B $, as 
it was done in the Einstein case, NDL assumes that this symmetry is maintained 
even after the introduction of nonlinearities. 

In a previous paper \cite{NDL}  the Feynman approach 
to nonlinear field theory of gravity that led to general relativity through 
an infinite series of self-interaction processes has been re-examined.

We extended the standard Feynman-Deser approach of field theoretical 
derivation of Einstein\rq s gravitational theory. It was then
possible to show how to obtain  a theory that incorporates a
great part of general relativity and can be interpreted in the 
standard geometrical way like GR, as far as the interaction of matter
to gravity is concerned. The most important particularity of the new theory 
concerns the gravity to gravity interaction. This theory satisfies all 
standard tests of gravity and lead to new predictions about the propagation
of gravitational waves. 
Since there is a large expectation that the detection of 
gravitational waves will occur in the near future, the question of which 
theory describes nature better will probably be settled soon.

The Lagrangian for the gravitational field in the NDL theory is given by:
\begin{equation} 
L = \frac{b^2}{\kappa} \left\{ \sqrt{1 - \frac{U}{b^{2}}} 
 - 1 \right\}, 
\label{Lag}
\end{equation}
where $U$ is defined by 
\begin{equation}
U\equiv A-B.
\label{U}
\end{equation}
The gravitational action is expressed as:
\begin{equation}
S = \int{\rm d}^{4}x\sqrt{-\gamma}\,L,
\label{action}
\end{equation}
where $\gamma$ is the determinant of the Minkowskian spacetime 
metric $\gamma_{\mu\nu}$ in an arbitrary coordinate system.
Taking the variation of the gravitational action (\ref{action}) with 
respect to the potential $\varphi_{\mu\nu}$, result in the 
following equations of motion:
\begin{equation}
\left[L_{U} F^{\lambda (\mu\nu)} \right]\mbox{}_{;\lambda} 
= -\frac{1}{2} T^{\mu\nu}
\label{eqmov}
\end{equation}
where $L_{U}$ represents the derivative of the Lagrangian 
with respect to the invariant $U$, and $T^{\mu\nu}$ is the energy-momentum 
tensor density of the matter contents.

Let us pause for a while in order to make contact with GR. For this, we 
express Eq. (\ref{eqmov}) under the form 
\begin{equation}
G^{\mbox{\tiny (L)}}_{\mu\nu} =  \chi_{\mu\nu} 
+ \frac{1}{2L_{U}} T_{\mu\nu} 
\label{3333}
\end{equation}
where the quantity $\chi_{\mu\nu}$ is provided by
\begin{equation}
\chi_{\mu\nu} \equiv \frac{L_{UU}}{L_{U}}U_{,\alpha} 
F^{\alpha}\mbox{}_{(\mu\nu)}.
\label{chi}
\end{equation}
One should compare this expression with the corresponding one 
(Eq. (\ref{112})) in GR.
It seems worth to remark that in the corresponding expression for GR in
place of $\chi_{\mu\nu}$ it appears precisely the {\it Gupta-Feynman
gravitational energy}.

Let us  make a short analysis of the wave propagation description in this
theory just for completeness. In what follows the symbol $[ J ]_{\Sigma}$ represents the 
discontinuity of the function $J$ through the surface $\Sigma$.

We set the following Hadamard\rq s \cite{Hadamard} discontinuity conditions :
\begin{equation}
[F_{\mu\nu\alpha}]_{\Sigma} = 0
\label{gw6}
\end{equation} 
and   
\begin{equation}
[F_{\mu\nu\alpha;\lambda}]_{\Sigma} = f_{\mu\nu\alpha} k_{\lambda},
\label{gw7}
\end{equation}
where $k_{\alpha}$ represents the wave vector normal to the surface of 
discontinuity $\Sigma$. The quantity $f_{\alpha\beta\gamma}$ has the same 
symmetries of $F_{\alpha\beta\gamma}$.
Taking the discontinuity of the equation of motion (\ref{eqmov}) we 
obtain\footnote{Note that this equation has a misprint in a original
formula as it appeared in Ref. \cite{NDL}.}:
\begin{equation}
f_{\mu(\alpha\beta)} k^{\mu} + 2 \frac{L_{UU}}
{L_{U}} ( \eta - \zeta ) F_{\mu(\alpha\beta)} k^{\mu} = 0
\label{gw8}
\end{equation}
in which the quantities $\eta$ and $\zeta$ are defined by
\begin{eqnarray} 
\eta &\equiv& F_{\alpha\beta\mu} f^{\alpha\beta\mu},
\nonumber\\
\zeta &\equiv& F_{\mu} f^{\mu}.
\end{eqnarray}
Considering the discontinuity relation and using the identities 
(\ref{d2}) and (\ref{d3}), after some algebraic manipulations it results: 
\begin{equation}
k^{\mu} k^{\nu} [ \gamma_{\mu\nu} + \Lambda_{\mu\nu} ] = 0,
\label{51}
\end{equation}
in which the quantity  $\Lambda_{\mu\nu}$ is written in terms of the 
gravitational field as:
\begin{equation} 
\Lambda_{\mu\nu} \equiv 2 \frac{L_{UU}}{L_{U}} 
[ {F_{\mu}}^{\alpha\beta} 
F_{\nu (\alpha\beta)} - F_{\mu} F_{\nu}  ]. 
\end{equation}
Note that the gravitational disturbances propagate in a 
modified geometry,
changing the background geometry $\gamma_{\mu\nu}$, into an effective 
one $g_{\mu\nu}$, which depends on the energy distribution of the field
$F_{\alpha\beta\mu}$. This fact shows that such a property stems from
the structural form of the Lagrangian\footnote{Indeed, we have shown 
recently that the same occurs for spin-1 field. See Ref. \cite{Novello}.}.

Differently from general relativity, in the NDL theory the 
characteristic surfaces of the 
gravitational waves propagate on the null cone of
an effective geometry distinct of that observed by all other forms 
of energy and matter. This result gives a possibility to
choose between these two theories just by observations of the 
gravitational waves. This is a challenge that is expected to be 
solved in the near future.

\section{Gravitational Quadrupole Emission}
\label{Quadrupole}

It has been observed that the orbital period 
$P_{b}$ of a binary system has a secular decrease. A plethora of effects 
may cause this, but the most important one is the emission of 
gravitational radiation \cite{Taylor79}.
The measurement of the change of the orbital period of this system due to 
radiation damping is in good agreement with 
the prediction of gravitational quadrupole emission of general relativity.
On the other hand, as it was pointed out by Eardley \cite{Eardley} and 
Will \cite{Will77}, the corresponding analysis of binary systems undertaken 
in the realm of most alternative
theories of gravity, predict gravitational dipole radiation. This is
a heavy drawback of these theories since the dipole contribution exceeds 
the corresponding general relativity quadrupole emission,
making this test a fundamental one.

In this section we will show that, in the NDL theory, the gravitational
radiation has a quadrupole origin and can be evaluated in a very analogous
way as in the GR theory.
We decided here to present this evaluation step by step in order
to compare with the standard evaluation formula from GR.

\subsection{Gravity Energy Momentum Tensor}
\label{Energy}

Since NDL is a field theory of gravity we can define
its corresponding gravitational energy momentum tensor through the
standard definition:
\begin{equation}
t_{\mu\nu} = \frac{2}{\sqrt {-\gamma}} \frac{\delta L
\sqrt{-\gamma}}{\delta \gamma^{\mu\nu}}.
\label{tmunu}
\end{equation}
From Lagrangian (\ref{Lag}) we obtain:
\begin{equation}
t_{\mu\nu} = -L \gamma_{\mu\nu} + 2L_{U} \left\{ 2
F_{\mu\alpha\beta}  {F_{\nu}}^{\alpha\beta} +  
F_{\alpha\beta\mu}  {F^{\alpha\beta}}_{\nu} -  F^{\alpha}
F_{\alpha(\mu\nu)} - F_{\mu} F_{\nu}  \right\}.
\label{tmunuF}
\end{equation}

Let us quote here that the corresponding Noether energy momentum tensor
\begin{equation}
N^{\alpha}\mbox{}_{\beta} = \varphi_{\mu\nu ,\beta}\frac{\partial L}
{\partial\varphi_{\mu\nu ,\alpha}} - \delta^{\alpha}\mbox{}_{\beta}L,
\label{Noether}
\end{equation}
reduces in our case to
\begin{equation}
N_{\mu\nu} = -L\gamma_{\mu\nu} - 2 L_{U}\varphi_{\alpha\beta ,\nu}
F_{\mu}\mbox{}^{\alpha\beta}.
\label{Noether2}
\end{equation}
The balance of energy between the gravitational field and its sources
takes the simple expression:
\begin{equation}
N^{\mu}\mbox{}_{\nu;\mu} = \frac{1}{2} 
T^{\alpha\beta}\varphi_{\alpha\beta;\nu}\, .
\end{equation}

\subsection{Energy Radiation}
\label{Radiation}

In the evaluation of the quadrupolar radiation  in
NDL theory, we will follow a similar procedure as the corresponding 
calculation in GR on this subject. We decided to do so in order to
exhibit the similitude and the different points concerning  both
theories.

The left hand side of Eq. (\ref{112}) and  (\ref{3333}) --- the linear part 
of both theories --- is gauge independent. We take, as usual,
the gauge condition\footnote{We remind the reader that all formulas are taken 
in the Minkowski background in a complete covariant way, that is, in
an arbitrary coordinate system.}:
\begin{equation}
\left(\varphi_{\alpha}\mbox{}^{\beta} - \frac{1}{2}\varphi\gamma_{\alpha}
\mbox{}^{\beta}\right)\mbox{}_{;\beta} = 0.
\label{gauge}
\end{equation}
Correspondingly we are thus led to define a new quantity
$h_{\alpha\beta}$ by setting :
\begin{equation}
h_{\alpha\beta} = \varphi_{\alpha\beta} - \frac{1}{2}\varphi\gamma_{\alpha
\beta}.
\label{defh}
\end{equation}
Using this into the equation of motion (Eq. (\ref{3333})), we obtain:
\begin{equation}
\Box h_{\mu\nu} = \chi_{\mu\nu} + \frac{1}{2L_{U}}\, T_{\mu\nu},
\label{eqh}
\end{equation}
in which the D'Alambertian operator is taken 
in the Minkowski background and $\chi_{\mu\nu}$ is given by Eq. (\ref{chi}).

Using the associate Green\rq s function the solution of 
$h_{\mu\nu}$ is
\begin{equation}
h_{\mu\nu}(\vec{x},t) = \int {\rm d}^{3} x' 
Z_{\mu\nu}(\vec{x}',t^{ret}=t-|\vec{x}-\vec{x}'|), 
\end{equation}
where we defined
\begin{equation}
Z_{\mu\nu}(\vec{x}',t^{ret}=t-|\vec{x}-\vec{x}'|) =
\frac{\left\{\chi_{\mu\nu} + \frac{1}{2L_{U}} 
T_{\mu\nu}\right\}_{ret}}{\left|\vec{x}-\vec{x'}\right|}.
\label{Z}
\end{equation}
We then expand this expression in the far region to obtain the series:
\begin{equation}
h_{\mu\nu}(\vec{x},t) = \frac{1}{R}\int {\rm d}^{3}x' Z^{ret}_{\mu\nu}
+ \frac{1}{R}\int {\rm d}^{3}x' Z^{ret}_{\mu\nu}\,\vec{x}.\vec{x}' 
+ ...
\end{equation}
where $R$ is the distance from the observer to the center of mass 
of the system.

\subsubsection{Quadrupolar Radiation}

Proceeding in analogy with GR (see Refs. \cite{Willbook,Landau} for more 
details) we obtain the expression of the first order for the quantity $h_{kl}$:
\begin{equation}
h_{kl} = \frac{2\mu}{R}\frac{\partial^{2}}{\partial t^{2}}(x_{k}x_{l})
+ O\left(\frac{1}{R}\right)^{2},
\label{hkl}
\end{equation}
where $\mu$ is the reduced mass of the system and $m$ is the total
mass $m = m_{1} + m_{2}$.  

Note that, using the gauge condition, we can obtain the temporal derivatives
of $h_{\mu 0}$  in terms of
the corresponding derivatives of the spatial components, that is:
\begin{eqnarray}
h^{k0}\mbox{}_{,0}& =& \hat{n}_{j} h^{jk}\mbox{}_{,0}
\label{h0k} \\
h^{00}\mbox{}_{,0}& =& \hat{n}_{j} \hat{n}_{k} h^{jk}\mbox{}_{,0}
\label{h00}
\end{eqnarray}
in which $\hat{n}_{k}$ is the unitary vector directed from the source to 
the observer,
\begin{equation}
\vec{n} = \frac{\vec{x}}{R}.
\end{equation}

In the case we are interested, that consists in the energy emitted by 
the system far from the source, only the gravitational contribution must be 
taken into account. This allows us to write:
\begin{equation}
\frac{{\rm d}E}{{\rm d}t} = -R^{2}\oint_{\Omega}{\rm d}\Omega\, 
t^{0j}\,\hat{n}_{j}.
\end{equation}

Expressing the exact energy momentum tensor (\ref{tmunuF}) in the new 
variables, results:
\begin{equation}
t_{\mu\nu} = -L\eta_{\mu\nu} + L_{U}\,Y_{\mu\nu}
\label{tmunuh}
\end{equation}
where $Y_{\mu\nu}$ is given by: 
\begin{eqnarray}
Y_{\mu\nu} &=& \left(
2 h_{\mu}\mbox{}^{\beta,\alpha}h_{\nu\beta,\alpha} - 
h_{\mu}\mbox{}^{\alpha,\beta}h_{\alpha\beta,\nu} -
h^{\beta\alpha}\mbox{}_{,\mu}h_{\nu\beta,\alpha} +
h^{\alpha\beta}\mbox{}_{,\mu}h_{\alpha\beta,\nu} \right.
\nonumber \\
&&- \left. h_{\mu\alpha ,\beta}h_{\nu}\mbox{}^{\beta ,\alpha} +
\frac{1}{2}h^{,\alpha}h_{\alpha\mu ,\nu} +
\frac{1}{2}h^{,\alpha}h_{\alpha\nu ,\mu} -
h^{,\alpha}h_{\mu\nu ,\alpha} -\frac{1}{2}h_{,\mu}h_{,\nu}\right).
\label{Q}
\end{eqnarray}
Note that since we are interested only on the post-Newtonian 
approximation, we can use the gauge condition (\ref{gauge}) in order to
simplify it. It is worth to remark that, in
the second order of $h_{\alpha\beta}$, the NDL gravitational energy
momentum tensor contains not only the associated gravitational
energy momentum tensor of the general relativity, e.g., the Landau
tensor, but extra terms.

%The $t_{0i}$ component can be written as
%\begin{equation}
%t_{0i} = t_{0i}^{\mbox{\tiny GR}} + outros\,termos
%\end{equation}
Using the gauge condition (up to a total divergence) results (see appendix
B for details):
\begin{eqnarray}
t_{0i} &=& -\frac{1}{32\pi}\left(-\frac{1}{2}\hat{n}_{a}\hat{n}_{b}
\hat{n}_{i}\hat{n}_{k}\hat{n}_{l}h^{ab}\mbox{}_{,0}h^{kl}\mbox{}_{,0} 
+2\hat{n}_{a}\hat{n}_{b}\hat{n}_{i}h^{ka}\mbox{}_{,0}h^{b}\mbox{}_{k,0}
\right.
\nonumber \\
&&\left.- \hat{n}_{a}\hat{n}_{b}\hat{n}_{i}h^{ab}\mbox{}_{,0}h^{k}\mbox{}_{k,0}
-\hat{n}_{i}h^{kl}\mbox{}_{,0}h_{kl,0} +
\frac{1}{2}\hat{n}_{i}h^{k}\mbox{}_{k,0}h^{l}\mbox{}_{l,0}\right).
\end{eqnarray}
Performing the angular integrations and averaging over several oscillations
of the sources yields\footnote{To perform the angular integration we use 
the following relations of the averages over the sphere:
\begin{equation}
\frac{1}{4\pi}\oint {\rm d}\Omega \hat{n}_{k}\hat{n}_{l} = 
\frac{1}{3}\delta_{kl}, \,\,\,\,\frac{1}{4\pi}\oint {\rm d}\Omega 
\hat{n}_{a}\hat{n}_{b}\hat{n}_{k}\hat{n}_{l} 
= \frac{1}{15}\left(\delta_{ab}\delta_{kl} + \delta_{ak}\delta_{bl}
+ \delta_{al}\delta_{bk}\right).
\end{equation}
}
\begin{equation}
\frac{{\rm d}E}{{\rm d}t} = - \frac{R^{2}}{8}\left< \frac{2}{5}
h^{kl}\mbox{}_{,0}h_{kl,0} 
- \frac{2}{15}h^{k}\mbox{}_{k,0}h^{l}\mbox{}_{l,0}\right>.
\label{epontohmedio}
\end{equation}

At this point it is convenient to introduce the standard traceless momentum of 
inertia tensor:
\begin{equation}
I_{kl} \equiv \mu\left(x_{k}x_{l} -\frac{1}{3}\delta_{kl}x^{2}\right).
\label{I}
\end{equation}
After some algebraic manipulations we obtain the expression for the 
rate of gravitational energy lost,  
\begin{equation}
\frac{{\rm d}E}{{\rm d}t} = 
- \frac{1}{5}\left< \stackrel{...}{I}\mbox{}^{kl}
\stackrel{...}{I}\mbox{}_{kl} \right>
\label{EpI}
\end{equation}
a formula which is precisely the same as one obtained in GR in this
order of approximation. Using the Newtonian result
\begin{equation}
\frac{{\rm d}v^{k}}{{\rm d}t} =-\frac{mx^{k}}{r^{3}}
\end{equation}
we arrive at the Peter-Mathews (PM) \cite{PM} expression:
\begin{equation}
\frac{{\rm d}E}{{\rm d}t} =-\frac{8}{15}
\left<\frac{\mu^{2}m^{2}}{r^{4}}\left(12v^{2}-11\dot{r}^{2}\right)\right>,
\label{PM}
\end{equation}
in which $v$ is the relative velocity and $\dot{r}$ is the temporal
derivative of the orbital separation $r$. 

By comparison of (\ref{PM}) and PM formula (see appendix C)  we obtain the 
values $\kappa_{1} = 1$, $\kappa_{2} = 1$. Note that as in GR there
is no dipole term for NDL theory.

\section{Conclusion}
\label{Conclusion}

In this paper we have continued the analysis of the NDL theory in what concerns
the observational tests of gravity. Here we treated
the emission of gravitational radiation by a binary system. We showed
that, in the post-Newtonian approximation, the results of this theory are
perfectly adjusted with the observational data, as well as in the general
relativity. 

In \cite{Taylor94} Taylor claims that ``the clock-comparison experiment for 
PSR $1913 + 16$  thus provides direct experimental proof that changes in 
gravity propagate at the speed of light, thereby creating a dissipative 
mechanism in an orbiting system. It necessarily follows that gravitational 
radiation exists and has a quadrupolar nature".
The second assertion (quadrupolar radiation) may be true independent from the 
first one (gravitational waves propagate at the  velocity of light) ---
which is the case in our theory.

From this remarks one can conclude that any theory which 
admits gravitational waves and gravitational radiation of quadrupolar nature 
is a good model. This does not prove that GR is correct, although 
it does prove that GR is a serious candidate to be the true theory of 
gravitational phenomena. From what we have shown in this paper,
NDL theory is a good candidate too.

\section{Acknowledgements}

The authors would like to thank the participants of the Pequeno Semin\'ario of 
Lafex/CBPF, particularly N.P. Neto and J.M. Salim for their comments. 
This work was supported by Conselho Nacional de Desenvolvimento Cient\'{\i}fico
e Tecnol\'ogico (CNPq) of Brazil.

\section*{Appendix A}

From the structural form of Lagrangians of the Born-Infeld \cite{Born} 
type -- as the one we have chosen to represent gravity processes -- it follows
that only even powers of the field variables $\varphi_{\mu\nu}$ appear 
in any polynomial-like expansion. This could be a 
drawback\footnote{In its quantum version this means 
that three-leg vertices are excluded.} of the
theory if, in the future, observation asks for the presence of the odd 
terms. There is an easy way to solve this problem leaving
the structural form of the theory intact. One has 
just to deal with a modification of the basic quantities 
by a re-definition of the field $F$ only through a change of 
$\varphi_{\mu\nu}$ by $\Psi_{\mu\nu}$, the nonlinear combination 
\begin{equation}
\Psi_{\mu\nu} = \varphi_{\mu\nu} - 
\varphi_{\mu\alpha} \,\varphi^{\alpha}\mbox{}_{\nu}.
\end{equation}

We leave the complete exam of this modification to a forthcoming paper.
Let us only inform here that, as far as the standard classical tests 
(PPN, cosmology and the binary pulsar) are concerned, both theories, in
the order of approximation dealt with, are undistinguishable.

\section*{Appendix B}

In this appendix we show explicitly the expansions of the terms
appearing in the expression (\ref{tmunuh}). Using the relations
between the components of the field $h_{\mu\nu}$, given by
(\ref{h0k}) and (\ref{h00}), results:  
\begin{eqnarray}
h_{0}\mbox{}^{\beta,\alpha}h_{ j\beta,\alpha}
&=& 
h_{00,0}h_{j0,0} + h_{k0,l}h_{jk,l} - h_{k0,0}h_{jk,0} - h_{00,l}h_{j0,l}
\nonumber\\
&=&
-\hat{n}_{a}\hat{n}_{b}\hat{n}_{k}h_{ab,0}h_{jk,0} 
-\hat{n}_{l}h_{kl,0}h_{jk,0}
+\hat{n}_{l}h_{kl,0}h_{jk,0}
+\hat{n}_{a}\hat{n}_{b}\hat{n}_{k}h_{ab,0}h_{jk,0} 
\nonumber\\
&=& 0;
\nonumber\\
h_{0}\mbox{}^{\alpha,\beta}h_{\alpha\beta, j}
& =& 
h_{00,0}h_{00,j} + h_{kl,j}h_{0k,l} - h_{k0,j}h_{0k,0} - h_{0l,j}h_{00,l}
\nonumber\\
&=&
-\hat{n}_{a}\hat{n}_{b}\hat{n}_{k}\hat{n}_{l}\hat{n}_{j}h_{ab,0}h_{kl,0} 
-\hat{n}_{b}\hat{n}_{l}\hat{n}_{j}h_{kl,0}h_{bk,0}
\nonumber\\
& & 
+\hat{n}_{l}\hat{n}_{b}\hat{n}_{j}h_{kl,0}h_{bk,0}
+\hat{n}_{a}\hat{n}_{b}\hat{n}_{k}\hat{n}_{l}\hat{n}_{j}h_{kl,0}h_{ab,0} 
\nonumber\\
&=& 0;
\nonumber\\
h^{\beta\alpha}\mbox{}_{,0}h_{ j\beta,\alpha} 
&=&
h_{00,0}h_{0j,0} + h_{kl,0}h_{jk,l} - h_{k0,0}h_{jk,0} - h_{0l,0}h_{j0,l}
\nonumber\\
&&
-\hat{n}_{a}\hat{n}_{b}\hat{n}_{k}h_{ab,0}h_{jk,0} 
-\hat{n}_{l}h_{kl,0}h_{jk,0}
+\hat{n}_{l}h_{kl,0}h_{jk,0}
+\hat{n}_{a}\hat{n}_{b}\hat{n}_{k}h_{ab,0}h_{jk,0} 
\nonumber\\
&=& 0.
\nonumber
\end{eqnarray}
In the same way we compute the other non vanishing terms:
\begin{eqnarray}
h^{\alpha\beta}\mbox{}_{,0}h_{ \alpha\beta, j} &=&
-\hat{n}_{a}\hat{n}_{b}\hat{n}_{k}\hat{n}_{l}\hat{n}_{j}h_{ab,0}h_{kl,0} 
+2\hat{n}_{a}\hat{n}_{b}\hat{n}_{j}h_{ka,0}h_{kb,0}
-\hat{n}_{j}h_{kl,0}h_{kl,0},
\nonumber\\
h_{,0}h_{, j} &=& -\hat{n}_{a}\hat{n}_{b}\hat{n}_{k}\hat{n}_{l}\hat{n}_{j}
h_{ab,0}h_{kl,0} +2\hat{n}_{a}\hat{n}_{b}\hat{n}_{j}h_{ab,0}h_{kk,0}
-\hat{n}_{j}h_{kk,0}h_{ll,0}.
\nonumber
\end{eqnarray}

\section*{Appendix C}

Just for completeness let us reproduce here the Peters-Mathews formula
for radiation emitted quoted in the text:
\begin{equation}
\dot{E} = -\left<\frac{\mu^2m^2}{r^4}\left[\frac{8}{15}\left(\kappa_1v^2
-\kappa_2\dot{r}^2\right) + \frac{1}{3}\kappa_D\sigma^2\right]\right>.
\label{GPM}
\end{equation} 
where the constants $\kappa_{i}$ depend on the each particular theory
for gravity phenomena.


\begin{thebibliography}{100}

\bibitem{Deser} 
S.Deser, J. Gen. Rel. Grav. {\bf 1}, 9, (1970).

\bibitem{GPP} 
L.P.Grischuck, A.N.Petrov and A.D.Popova, Commun. 
Math. Phys. {\bf 94}, 379 (1984).

\bibitem{Gup54}
S.N. Gupta, Phys. Rev. {\bf 96}, 1683 (1954).

\bibitem{Krai55}
R. H. Kraichnan, Phys. Rev. {\bf 98}, 1118 (1955).

\bibitem{Feynman} 
R.P.Feynman, in {\it Feynman Lectures On Gravitation}, (Addison-Wesley
Pub. Company, Massachusetts, 1995).

\bibitem{Fierz}
M. Fierz and W. Pauli, Proc. Roy. Soc. {\bf 173A}, 211 (1939).

\bibitem{Willbook}
C.M. Will, in {\it Theory and Experiment in Gravitational Physics},
(Cambridge University Press 1981, 1983).

\bibitem{NDL}
M. Novello, V.A. De Lorenci and L.R. de Freitas,  
Annals of Phys. {\bf 254}, $n^o_.$ 1, 83 (1997).

\bibitem{Hadamard}
J. Hadamard, in {\it Lectures  on Cauchy\rq s  Problem},
(Yale University Press 1923); Dover reprint (1952).

\bibitem{Novello}
M. Novello, V.A. De Lorenci and E. Elbaz, Report $n^o_.$ gr-qc/9702054.

\bibitem{Taylor79}
J.H. Taylor, L.A. Fowler and M. McCulloch, Nature {\bf 277}, 437 (1979).

\bibitem{Eardley}
D.M. Eardley, Astrophys. J. Lett. {\bf 196}, L59 (1975).

\bibitem{Will77}
C.M. Will, Astrophys. J. {\bf 214}, 826 (1977).

\bibitem{Landau}
L.D. Landau and E.M. Lifshitz, {\it in The Classical Theory of Fields},
(Addison-Wesley, Reading. Mass 1962); C.W. Misner, D.S. Thorne and J.A.
Wheeler, in {\it Gravitation}, (Freeman, San Francisco, 1973).

\bibitem{PM}
P.C. Peters and J. Mathews. Phys. Rev. {\bf 131}, 435 (1963).

\bibitem{Taylor94}
J. H. Taylor, Rev. Mod. Phys. {\bf 66}, 711 (1994).

\bibitem{Born} 
M.Born, Nature, {\bf 132}, 282 (1933); 
ibid, Proc. Roy. Soc. {\bf A 143}, 410 (1934); 
M.Born and L.Infeld, Nature {\bf 133}, 63 (1934).

\end{thebibliography}
\end{document}